\documentclass[aps,prl,twocolumn,superscriptaddress]{revtex4}
\usepackage{graphicx}
\input epsf
\usepackage{amsmath,amssymb}

\begin{document}

\title{Micron size superconducting quantum interference devices of lead (Pb)}
\author{Sagar Paul}
\affiliation{Department of Physics, Indian Institute of Technology Kanpur, Kanpur 208016, India}
\author{Sourav Biswas}
\affiliation{Department of Physics, Indian Institute of Technology Kanpur, Kanpur 208016, India}
\author{Anjan K. Gupta}
\email[]{anjankg@iitk.ac.in}
\affiliation{Department of Physics, Indian Institute of Technology Kanpur, Kanpur 208016, India}
\date{\today}

\begin{abstract}

Micron size superconducting quantum interference devices ($\mu$-SQUID) of lead (Pb), for probing nano-magnetism, were fabricated and characterized. In order to get continuous Pb films with small grain size, Pb was thermally evaporated on a liquid nitrogen cooled Si substrate. Pb was sandwiched between two thin Cr layers for improved adhesion and protection. The SQUID pattern was made by e-beam lithography with Pb  lift-off after deposition. The current-voltage characteristics of these devices show a critical current, which exhibits the expected SQUID oscillations with magnetic field, and two re-trapping currents. As a result these devices have hysteresis at low temperatures, which disappears just below the critical temperature.

\end{abstract}

\maketitle

\section{Introduction}
A superconducting weak link (WL) based micron- or nanometer- size superconducting quantum interference device ($\mu$- or nano-SQUID) is the most suitable probe for nano-magnetism with demonstrated applications in single nano-particle magnetometry \cite{mag-nano-ptcle} and in scanning $\mu$- or nano-SQUIDs \cite{scan-SQUID,zeldov-SOT}. A superconducting WL (such as a constriction) between two superconductors behaves like a `Josephson junction' \cite{likharev}, and thus it allows the flow of coherent supercurrent, up to a magnitude $I_C$. In a SQUID, consisting of two parallel WLs in a loop, $I_C$ oscillates with the flux threading through the loop with a periodicity ${\Phi}_0$ = 2.07$\times$10$^{-15}$ Wb \cite{tinkham book}. Thus SQUIDs have been used to detect the magnetic flux with sensitivity better than $10^{-7}\Phi_0$ \cite{SQUID-handbook}. Present research on $\mu$-SQUID can be broadly divided into three categories, namely, 1) controlling hysteresis \cite{scocpol,herve,hazra-prb,nikhil prl,hazra-PRA,Blois-proximity-SQUID,Nikhil-SuST} and improving sensitivity, 2) making smaller loop-area SQUIDs for improved magnetic coupling with nano-particles and for better spatial resolution in scanning-SQUIDs \cite{zeldov-SOT,Blois-proximity-SQUID,lam-apl}, and 3) improving materials and processes either for ease of fabrication or for surpassing the working limits, for instance temperature \cite{hi-tc-SQUIDs} and magnetic field \cite{diamond-SQUID}, of the existing devices.

A number of elemental superconducting materials have been used for fabricating $\mu$-SQUIDs with the most common being Nb. However, good quality Nb films require deposition in UHV conditions and Nb superconductivity readily degrades in the lift-off process due to the presence of soft polymer in the vicinity. Special lift-off methods for Nb involving multi-layer lithography have been successfully demonstrated \cite{Nb-lift-off}. Among various elemental superconductors, lead (Pb) thin films, having $T_C$ same as bulk, are easily deposited in non-UHV systems and also its superconductivity is not expected to degrade in presence of polymer patterns as the evaporation temperature is not so high. However, being a low melting point metal, it can easily diffuse \cite{pbdiffusivity} and make films with large disconnected grains \cite{pbdepo1,pbdepo2,pbpld}. Although Pb conventional tunnel junctions were successfully made earlier \cite{Pb-SQUID} but these were found to have micro-shorts problem due to thermal-cycling induced stress. Such problems are unlikely to affect WL devices as the recent work on Pb nano-SQUID on a tip shows \cite{zeldov-SOT}.

In this paper, we report on SC WL devices and $\mu$-SQUIDs of lead (Pb), with $T_C$ = 7.2 K. The devices were fabricated using electron beam lithography followed by Cr (5nm)/Pb (40nm)/Cr (8nm) deposition on liquid nitrogen cooled Si substrate and lift-off. Pb was sandwiched between two Cr layers for improved adhesion and protection. For ease of comparison and understanding we used similar design as our previously studied Nb devices \cite{nikhil prl}. Electrical transport measurements down to 1.3 K on a WL device and a $\mu$-SQUID show the expected behavior. The ambient-life of these devices, due to oxidation, is found to be between 2 and 3 months, which is long enough to be useful.
\begin{figure}
\begin{center}
\includegraphics[width=6.5cm]{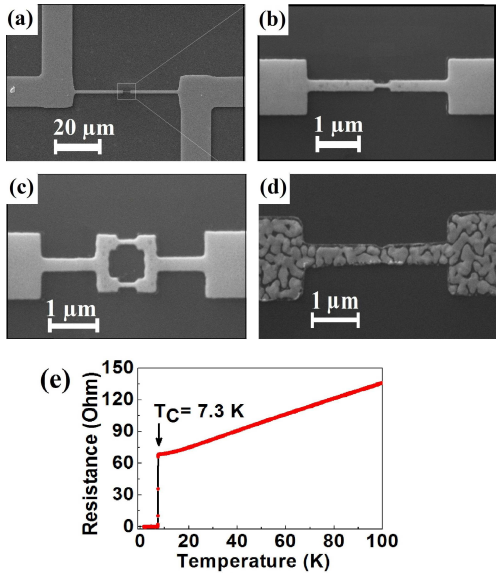}
\caption{(a) Electron micrograph of a WL device having WL width 80 nm and length 350 nm; (b) shows the enlarged view of the same. (c) Electron micrograph of a $\mu$-SQUID with same WL dimensions. (d) electron micrograph of a failed device due to room temperature Pb deposition showing large and disconnected grains. (e) shows the resistance of the WL device as a function of temperature with SC transition at $T_C= $ 7.3 K.}
\label{fig1}
\end{center}
\end{figure}

\section{Device Fabrication}
The device fabrication involved three steps, namely, 1) Electron beam lithography (EBL) of PMMA (M$_W$ = 950 k) resist on a Si substrate in the desired pattern, 2) resistive thermal evaporation of Cr/Pb/Cr tri-layer on the patterned substrate 3) lift-off using acetone. For EBL we coated 4$\%$ of 950 k PMMA (in anisole) as resist on Si substrate at 4000 rpm giving a thickness of $\sim$ 200 nm. The PMMA was cured and exposed with 20 kV and 100 pA e-beam in a Zeiss electron microscope (Model: MA15 with LaB$_6$ filament) interfaced with a Nabity writer. The optimal exposure dose was found to be 170 $\mu$C/cm$^2$ after experimenting over 130 to 200 $\mu$C/cm$^2$ range with Cr/Au films.

For successful device fabrication, continuous Pb films with grain-size much smaller than the smallest-size features of the pattern are essential. Pb forms disconnected large grains or islands (see fig.\ref{fig1}d) if the deposition is done on substrate at room temperature \cite{pbdepo1,pbdepo2,pbpld}. Thus we mounted the substrate on a liquid nitrogen cooled copper surface with good thermalization \cite{pbdiffusivity} inside the evaporation chamber having base pressure below $10^{-6}$ mbar. The Pb source was degassed a few times before deposition. A good thermal contact between cold copper surface and the substrate was ensured by polishing the copper surface as otherwise the Pb grain-size was not sufficiently small.

5 nm Cr was deposited before 40 nm Pb for its good adhesion with Si substrate followed by another 8nm Cr for preventing Pb oxidation. The lift-off was carried out using acetone for about 1 hour. Figure \ref{fig1} shows the SEM images of a WL device and a $\mu$-SQUID thus fabricated. The design of the devices between the voltage and current leads is similar to our previously studied Nb devices \cite{nikhil prl} with contact leads and pads having slight variation.

\section{Transport measurements and analysis}
Electrical transport measurements down to 1.3 K temperature of these devices were done in a closed cycle He-refrigerator in four probe configuration. Indium press contacts were used to make connections with Pb contact pads. The devices were characterized using homemade voltage controlled current source and voltage amplifier interfaced with a data acquisition card. A LabView program was used for the measurements.

The resistance of the WL device as a function of temperature in fig.\ref{fig1}(e) shows a single superconducting transition at $T_C= $ 7.3 K. This is unlike the Nb devices where different portions of the device showed slightly different T$_C$ values \cite{nikhil prl}. The residual resistivity ratio (RRR) defined as $\rho (300K)/\rho(10K)$ was found to be 4. Using the lateral dimensions of Pb film between the voltage leads, the sheet resistance just above T$_C$ is found to be 1.25 $\Omega$. Ignoring the Cr contribution to conduction and using 40nm Pb thickness its resistivity just above T$_C$ works out as 5 $\mu\Omega.cm$. The T$_C$ value same as bulk Pb and reasonably large RRR value establish good film quality. We note that the presence of Cr does not affect the T$_C$ value of Pb as opposed to some of the other metals that have been tried for protecting Pb and for reducing thermal stresses in Pb devices \cite{Pb-th-stress}. In fact Cr itself superconducts in thin film form although with a T$_C$ limited to below 3 K \cite{Cr-SC}.

\begin{figure}
	\begin{center}
		\includegraphics[width=7.0cm]{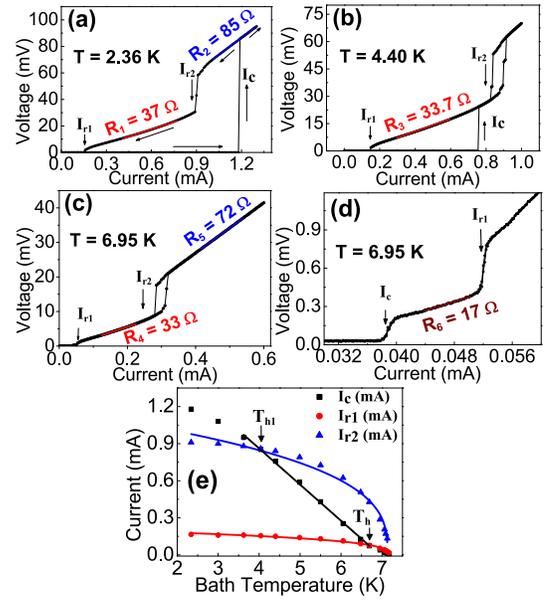}
		\caption{(a)-(c) IVCs of the WL device showing the voltage jumps at $I_C$, $I_{r1}$ and $I_{r2}$ at different bath temperatures. (d) shows the low bias-current portion of the IVC shown in (c) in order to resolve the I$_C$ transition for this non-hysteretic IVC. (e) shows the I$_C$, I$_{r1}$ and I$_{r2}$ variation with temperature. Here the continuous lines are fits given by $I_C$ = 0.293(6.94-T$_b$) mA, I$_{r1}$ = 0.1(7.14-T$_b$)$^{0.34}$ mA and I$_{r2}$ = 0.58(7.14-T$_b$)$^{0.34}$ mA.}
		\label{fig2}
	\end{center}
\end{figure}

Figure \ref{fig2} shows the temperature evolution of current voltage characteristics (IVC) of the WL device, which are hysteretic and show a critical current (I$_C$) and two re-trapping currents, I$_{r1}$ and I$_{r2}$. Such hysteresis is common in the DC IVCs of the Josephson junction devices. The hysteresis in Josephson junctions with tunnel barrier is attributed to the large junction capacitance \cite{tinkham book} whereas the hysteresis in WL devices occurs due to resistive Joule heating \cite{scocpol,tinkham nano wire} as evidenced from direct correlation between hysteresis and local temperature \cite{herve}. In WL devices hysteresis commonly exists till a crossover temperature, $T_h$($<T_C$) \cite{hazra-prb} as $I_C$ and $I_r$, controlled by different physics, have different temperature dependence. The re-trapping currents $I_{r1}$ and $I_{r2}$, respectively, seen here arise from the thermal instabilities of normal-superconductor (NS) interface \cite{nikhil prl} in different portions of these devices.

The IVC slope in fig.\ref{fig2}(c) above I$_{r2}$ is R$_5=$ 72$\Omega$ at 6.95 K, which is close to the resistance just above T$_C$ [see fig.\ref{fig1}(e)] indicating that whole of the device portion between voltage leads is resistive above I$_{r2}$. At low temperatures, a higher slope (R$_2=$ 85$\Omega$) arises from significant temperature rise due to higher current values leading to higher resistivity. Similarly, the slopes of the IVC branch between I$_{r2}$ and I$_{r1}$ (i.e. R$_1$, R$_3$ and R$_4$) are consistent with the expected resistance, i.e. 29 $\Omega$, of the narrow leads and the WL combined. In the non-hysteretic regime the value of the IVC slope between I$_C$ and I$_{r1}$, i.e. R$_6$ = 17 $\Omega$ [see fig.\ref{fig2}(d)], is larger than the expected WL resistance of 6 $\Omega$. This indicates spread of the resistive hot-spot into the adjoining narrow leads in this regime. It is likely that there is a third re-trapping current in this device related to the shift of NS interface from inside WL to the adjoining narrow leads, which is merged with I$_C$ in this limited temperature range.

As seen in fig. \ref{fig2}(e), I$_{r2}$ and I$_{r1}$ cross I$_C$ at T$_{h1} = $ 4 K and T$_h = $ 6.7 K, respectively. So this WL device is hysteretic over most of the superconducting temperature range except between 6.7 K and 7.3 K. It is interesting to note that the transition at I$_{r2}$ [see fig.\ref{fig2}(b) and (d)], due to instability of NS interface in the wide leads, shows visible hysteresis even above T$_{h1}$. This could arise due to heating induced reduction in critical current density making the wide leads superconduct only when the bias current becomes somewhat lower than $I_{r2}$.

\begin{figure}
\begin{center}
\centering
\includegraphics[width=6.3cm]{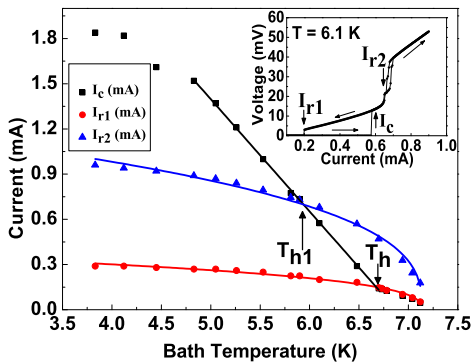}
\caption{I$_C$, I$_{r1}$ and I$_{r2}$ variation with temperature for the SQUID device. The continuous lines are fits given by I$_C$ = 0.745(6.876-T$_b$) mA, I$_{r1}$ = 0.19(7.13-T$_b$)$^{0.35}$ mA and I$_{r2}$ = 0.65(7.14-T$_b$)$^{0.36}$ mA. Inset: An IVC of the SQUID at 6.1 K.}
\label{fig3}
\end{center}
\end{figure}

The wide leads in our device are much longer than $l_{th}$=$\sqrt{\kappa t/\alpha}$ \cite{nikhil prl} representing a length scale over which the temperature relaxes. Here $\kappa$ is the thermal conductivity, $t$ is film thickness and $\alpha$ is the interface heat loss coefficient. Thus if the lead length $l\gg$ $l_{th}$ we can assume the leads to be semi-infinite for finding the re-trapping (or thermal-instability) current as $I_{r}=w\sqrt{2 \alpha (T_{c}-T_b) / R_{\Box}}$ \cite{nikhil prl} with $w$ as the lead width and $R_{\Box}$ as the sheet resistance. We use the Wiedemann-Franz law, i.e. $\kappa = LT/\rho$ with $L$ = 2.44x10$^{-8}$ W$\Omega$/K$^2$ as the Lorenz number, $T = T_c$ = 7.3 K and $\rho$ = 5 $\mu\Omega.cm$ to get $\kappa$=3.56 W/m.K. We use $I_{r2}$ value at low temperature in the above expression for re-trapping current to get $\alpha$ = 12.4 $W/cm^2.K$ and this in turn gives $l_{th}$ as about 1 $\mu m$ which is much smaller than the length of the wide leads.

We have also fitted the variation of $I_C$, $I_{r1}$ and $I_{r2}$ of the WL device with bath temperature as shown in fig.\ref{fig2}(e). We fit a linear variation of I$_C$ with T$_b$ for $T_h>T_b>3.5$ K and find $I_C$ = 0.29(6.9-T$_b$) mA. With 6 $\Omega$ estimated WL resistance we get a slope of $I_C$R$_N$ Vs T$_b$ as 180 $\mu$V/K, which is smaller than the general expected value of 635 $\mu$V/K for WLs \cite{likharev}. We note that our WL length is quite large ($\sim$350 nm) and it exceeds the Pb coherence length (about 80 nm, in clean limit) except for temperatures close to T$_C$. The two re-trapping currents are fitted to I$_{r1}$ = 0.1(7.14-T$_b$)$^{0.34}$ mA and I$_{r2}$ = 0.58(7.14-T$_b$)$^{0.34}$ mA. The fitting exponent is closer to 1/3 instead of 1/2 expected for the temperature independent $\kappa$. This exponent is close to 1/3 in the thermal models for WL with very wide electrodes and with temperature dependent $\kappa$ \cite{hazra-prb}.
\begin{figure}
\begin{center}
\includegraphics[width=6.3cm]{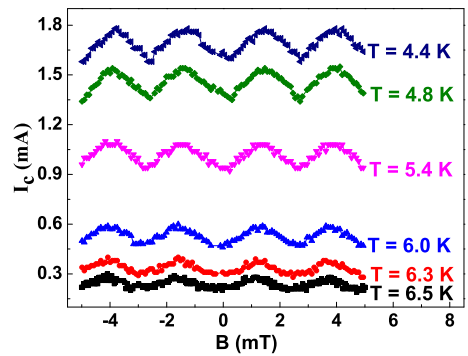}
\caption{I$_C$ oscillations with magnetic field for the SQUID at different temperatures in hysteretic regime.}
\label{fig4}
\end{center}
\end{figure}

Fig. \ref{fig3} shows an IVC and temperature variation of I$_C$, I$_{r1}$ and I$_{r2}$ for the $\mu$-SQUID [see fig.\ref{fig1}(c)]. The IVCs  have a similar behavior as that of the WL discussed earlier. The magnitude of I$_{r2}$ here is higher than that of the WL device as the design and dimensions of the region in between the wide leads differ [see fig.\ref{fig1}(b) and (d)]. Fig. \ref{fig4} shows the I$_C$ oscillations with magnetic field at different temperatures. From the period of oscillation and flux quantum value the expected SQUID loop area is 0.83 $\mu m^2$ which is close to the area (0.81 $\mu m^2$) inferred from electron micrographs. We did not see voltage oscillations with magnetic field in the non-hysteretic regime, which is rather narrow in temperature. As discussed earlier, in the non-hysteretic regime, the IVC slope for the WL device was found to be significantly higher than the expected WL resistance value indicating heating beyond WL. The same is found to be true for the studied $\mu$-SQUID. This is suspected to be the reason for not seeing any voltage oscillations in the non-hysteretic regime. The WL together with a significant portion of the narrow leads become normal across I$_C$.

\section{Discussions and Conclusions}

We have also characterized these devices after storage in ambient conditions and found the shelf-life to be between two and three months, which is long enough to be useful for probing nano-magnetism. On the other hand the fabrication is not so involved and thus such devices can be easily made in less than two days time. In order to use these for probing nano-magnetism, some more optimizations and studies are needed. In particular, we would like to reduce I$_C$ values by using WLs of smaller cross-section area and also of reduced length to get close to ideal behavior. This should also help in getting wider temperature range of hysteresis free operation, which can also be achieved by putting a normal metal shunt in the close vicinity \cite{Nikhil-SuST}.

In conclusion, we have presented a complete process for fabricating Pb based superconducting weak link devices and $\mu$-SQUIDs using electron lithography and thermal evaporation technique. The devices show expected IVCs and I$_C$ oscillations with magnetic field.

\section{Acknowledgements}
This work has been financed by the CSIR and DST of the Govt. of India.

\end{document}